\documentclass[12pt, a4paper]{article}
\usepackage[dvips]{graphicx}
\usepackage[hang]{caption2}

\begin{document}

%%%%%%%%%%%%% User definition %%%%%%%%%%%%%%%%
\newcommand{\PRI}{\mathrm{PRI}}
%%%%%%%%%%%%%%%%%%%%%%%%%%%%%%%%%%%%%%%%%%%%%%

{\large\bfseries\noindent Algorithms of Two-Level
Parallelization for DSMC of Unsteady Flows in Molecular Gasdynamics
\medskip

\noindent A.V. Bogdanov, N.Yu. Bykov, I.A. Grishin,
Gr.O. Khanlarov, G.A. Lukianov and V.V. Zakharov
}
\medskip
\begin{flushright}
\footnotesize submitted for publication to the conference HPCN'99
\end{flushright}

{\large\bfseries\noindent Abstract\\}
The general scheme of two-level parallelization (TLP) for direct 
simulation Monte Carlo of unsteady gas flows on shared memory
multiprocessor computers 
has been described. The high efficient algorithm of parallel independent runs 
is used on the first level. The data parallelization is employed for the second 
one. 

Two versions of TLP algorithm are elaborated with static and dynamic load 
balancing. The method of dynamic processor reallocation is used for dynamic 
load balancing. 

Two gasdynamic unsteady problems were used to study
speedup and efficiency of the algorithms. The conditions of efficient 
application field for algorithms have been determined.

\copyright Institute for High Performance Computing and Data Bases
\vspace{1.0cm}

\newpage

\hyphenpenalty=5000
\section{Introduction}
\subsection[Direct Simulation Monte Carlo Method and Sequential Algorithm
in Unsteady Molecular Gasdynamics]%
{Direct Simulation Monte Carlo Method and Sequential\\
Algorithm in Unsteady Molecular Gasdynamics}

The Direct Simulation Monte Carlo (DSMC) is the simulation of  real gas 
flows  with various physical processes by means of huge number of modeling 
particles~\cite{bird}, each of which is a typical representative of great 
number of real gas particles (molecules, atoms, etc.).
The DSMC method conditionally divides the
continuous process of particles movement and collisions into two 
consecutive stages (motion and collision process) at each time step 
$\Delta t$. The particle parameters (coordinates, velocity)
are stored in the computer's memory.
To get information about the flow field the computational domain has to be 
divided into cells. The results of simulation are averaged particles
parameters in cells.
\begin{figure}[hb]
\centering
\includegraphics[scale=0.7]{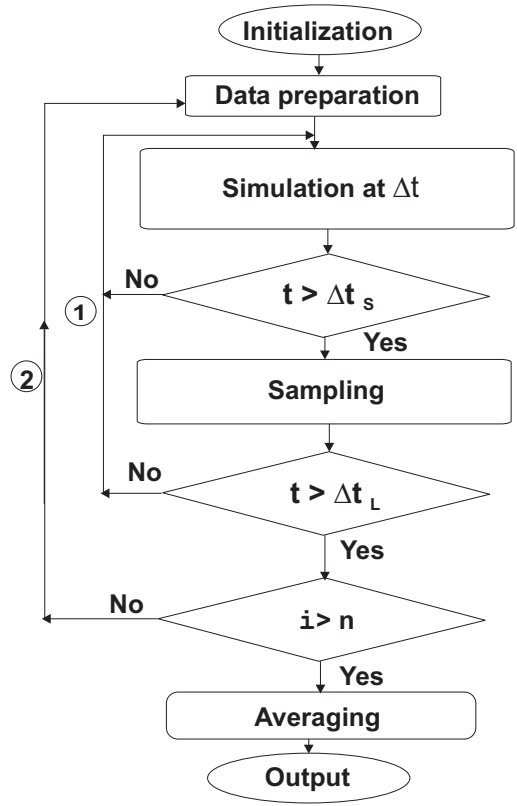}
\caption{General flowchart of sequential algorithm for DSMC
of unsteady flows.
$\Delta t$ --- time step,
$\Delta t_s$ --- interval between samples,
$\Delta t_L$ --- total time of a single run,
$t$ --- current time,
$n$ --- number of runs,
$i$ --- iteration number.}
\label{f:seq_mc}
\end{figure}

The finite memory size and computer performance make restrictions to the 
total number of modeling particles and cells. Macroscopic gas parameters 
determined by particles parameters in cells at the current time step are the 
result of simulation. Fluctuations of averaged gas parameters at single time 
step can be rather high owing to relatively small number of particles in cells. 
So, when solving steady gasdynamic problems, we have to increase the time
interval of averaging (the sample size) after steady state is achieved
in order to reduce statistical error down to the required level.
The averaging time step $\Delta t_{av}$ has to be much
greater than the time step $\Delta t$ ($\Delta t_{av}\gg\Delta t$).

For DSMC of unsteady flows the value of averaging time step 
$\Delta t_{av}$ for a given problem and at the current time $t$ has to meet
the following requirement: $\Delta t_{av}\ll\min t_H(x,y,z,t)$, where 
$t_H$ --- is the characteristic time of flow parameters variation. The choice 
of the value of $t_H$ is determined by particular problem~\cite{bykov,bogd}. 
In order to meet the condition for the averaging interval we have to carry out 
enough number of statistically independent calculations (runs) $n$ to get the 
required sample size. This leads to the increase of the total calculation time 
which is proportional to $n$ in the case of sequential DSMC algorithm.

The general flowchart of classic sequential algorithm~\cite{bird}
is depicted in the fig.~\ref{f:seq_mc}.
The algorithm of DSMC of unsteady flows consists of two basic loops. In 
the first (inner) loop the single run of unsteady process is executed.
First, we generate particles at input boundaries of the domain
(subroutine \verb+Generation+). Then we carry out
simulation of particle movement, surface interaction (subroutine 
\verb+Motion+) and collision process (subroutine \verb+Interaction+) for 
determined number of time steps $\Delta t$. The sampling (subroutine 
\verb+Sampling+)  of flow macroparameters in cells is carried out at a given 
moment of unsteady process. The inner loop itself is divided into two 
successive steps. At the first step we sequentially carry out simulation for 
each of $N_p$ particles independently. A special readdressing array is formed 
-- subroutines \verb+Enumeration+, \verb+Indexing+ -- (it determines the 
mutual correspondence of particles and cells) after the first step. We have to 
know the location of all particles in order to fill that array. At the second step 
we carry out the simulation for each of  $N_c$ cells independently. For 
$t>\Delta t_s$ we accumulate statistical data of flow parameters in cells.

The second (outer) loop repeats unsteady runs $n$ times to get the desired 
sample size. Each run is executed independently from the previous ones. To 
make separate unsteady runs independent we have to shift random number 
generator (\verb+RNG+).

For each unsteady run three basic arrays (\verb+P+, \verb+LCR+, \verb+C+) 
are required. The array~\verb+P+ is used for storing information about 
particles. The array \verb+LCR+ is the readdressing array. The dimensions of 
these arrays are proportional to the total number of particles. 
The array~\verb+C+ stores information about cells and macroparameters. The 
dimension of this array is proportional to the total number of cells of a
computational grid.
The DSMC method requires several additional arrays which reserve
much smaller memory size. The particles which abandon the domain
are removed from the array~\verb+P+, whereas the new generated
particles are inserted into the one.
Since the particles move from one cell to another we have to
rearrange the array \verb+LCR+ and update the array~\verb+C+.
These procedures are performed at each time step $\Delta t$.

\subsection{Parallelization methods for DSMC of gas flows}
The feasibility of parallelization and the efficiency of parallel algorithms 
are determined both by the structure of modeling process and by the 
architecture and characteristics of a computer (number of processors, memory 
size, etc.).

The development of any parallel algorithm starts with the decomposition of 
a general problem. The whole task is divided into a series of independent or 
slightly dependent sub-tasks which are solved parallel. For direct simulation 
of gas flows there are different decomposition strategies depending on goals 
of modeling and flow nature. The development of parallel algorithms for 
DSMC started not long ago (about 10 years ago). At the present time the 
common classification of principal types of parallel algorithms has not been 
formed yet. However, one can point out several approaches to parallelize the 
DSMC, the efficiency of which is proved by the practice of their usage. Let 
us conditionally single out four types of parallel algorithms of DSMC.

The first type is the parallelization by coarse-grained independent sub-tasks. 
This method has been realized in \cite{korolev}--\cite{bykov} for 
parallelization of  DSMC of unsteady problems. The algorithm consists in 
the reiteration of statistically independent modeling procedures (runs) of a 
given flow on several processors.

The second type is the spatial decomposition of a computational domain. 
The calculation in each of regions are single sub-tasks which are solved 
parallel. Each processor performs calculations for particles and cells in its 
own region. The transfer of particles accompanies with data exchange 
between processors. Therefore, these sub-tasks are not independent.

This method of parallelization is the most widespread at the present for 
parallel DSMC of both steady and unsteady flows \cite{furlani}--\cite{robinson2}.
The main advantage of this approach is the reduction of 
memory size required by each processor. This method can be carried out on 
computers with both local and shared memory. The method has drawback 
for increasing number of processors: the increase of the number of 
connections between regions and the increase of relative amount of data to 
exchange between regions. The essential condition of high efficiency of this 
method is the ensuring of uniform load balancing and minimization of data 
exchange. One can use static and dynamic load balancing to make good load 
balancing. The modern parallel algorithms of this type usually employ 
dynamic load balancing.

The third type is the algorithmic decomposition. This type of parallel 
algorithms consists in the execution of different parts of the same procedures 
on different processors. For realization of these algorithms it is necessary to 
use a computer with architecture which is adequate to a given algorithm. 
The examples of this type of algorithm is the data parallelization 
\cite{oh,grishin}.

The fourth type is the combined decomposition which includes all types 
considered precedingly. The decomposition of computational domain with 
data parallelization are carried out in~\cite{oh}. In this paper we shall 
consider two-level algorithms which include methods of first and third type.

\subsection[Algorithm of Parallel Statistically Independent Runs]%
{Algorithm of Parallel Statistically\\ Independent Runs (PSIR)~\cite{bykov}}

\begin{figure}[hb]
\centering
\includegraphics[scale=0.6]{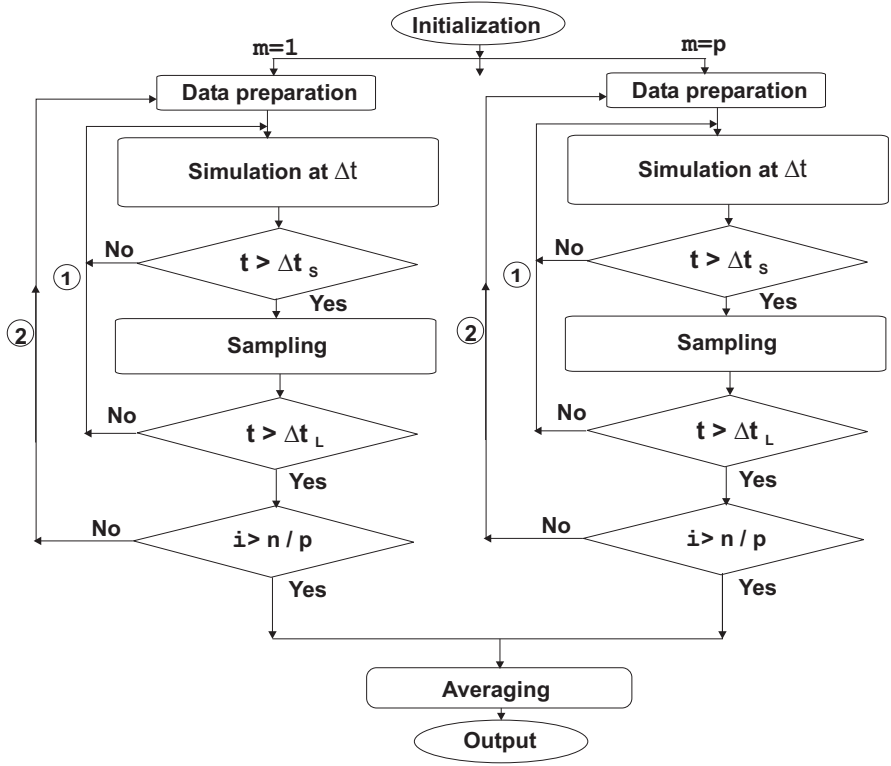}
\caption{General flowchart of PSIR algorithm;
$m$ --- processor ID,
$p$ --- number of processors.}
\label{f:psir_g}
\end{figure}

The statistical independence of  single runs make it possible to execute them 
parallel. The general flowchart of the PSIR algorithm is depicted
in the fig.~\ref{f:psir_g}. The implementation of this approach on a
multiprocessor computer leads to the decrease of the number of iterations of 
the outer loop for every single processor ($n/p$ --- the number of iterations 
for the $p$-processor computer). The data exchange between processors 
goes after all calculation are finished. Only one processor sequentially 
analyzes the results after data exchange. The range of efficient 
application field for this algorithm is $p\leq n$.
The value of $n$ has to be multiply by $p$ to get
optimal speedup and efficiency.

\begin{figure}
\centering
\includegraphics[scale=0.5]{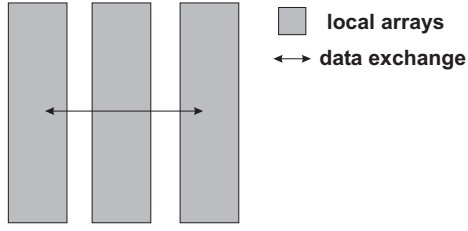}
\caption{Scheme of memory usage for PSIR algorithm
(the case of three processors).}
\label{f:psir_mem}
\end{figure}

All arrays (\verb+P, LCR, C+, etc.) are stored locally for each run. This 
algorithm can be realized on computers with any type of memory (shared or 
local). The message passing is used to perform data exchange on
computers with local memory. The scheme of memory usage is presented
in the fig.~\ref{f:psir_mem}.
The required memory size for this algorithm is proportional to $p$.

The speedup $S_p$ and the efficiency $E_p$ of parallel algorithm with a 
parallel fraction of computational work $\alpha$ for the computer with $p$
processors are as follows~\cite{ortega}:
\begin{equation}
S_p(p, \alpha)=\frac{T_1}{T_p},
\end{equation}
\begin{equation}
E_p(p, \alpha)=\frac{S_p}{p},
\end{equation}
where $T_1$ --- the execution time of the sequential algorithm,
$T_p$ --- the execution time of a given parallel algorithm on the computer 
with $p$ processors ($p$ --- number of reserved processors).
In this paper we use a model of computational process which
assumes that there is some parallel fraction $\alpha$
of total calculations and sequential fraction $(1-\alpha)$. 
The parallel and sequential calculations
are not coincided.
\begin{figure}[!hb]
\centering
\begin{minipage}[c]{.5\textwidth}
 \centering
  \includegraphics[width=\textwidth]{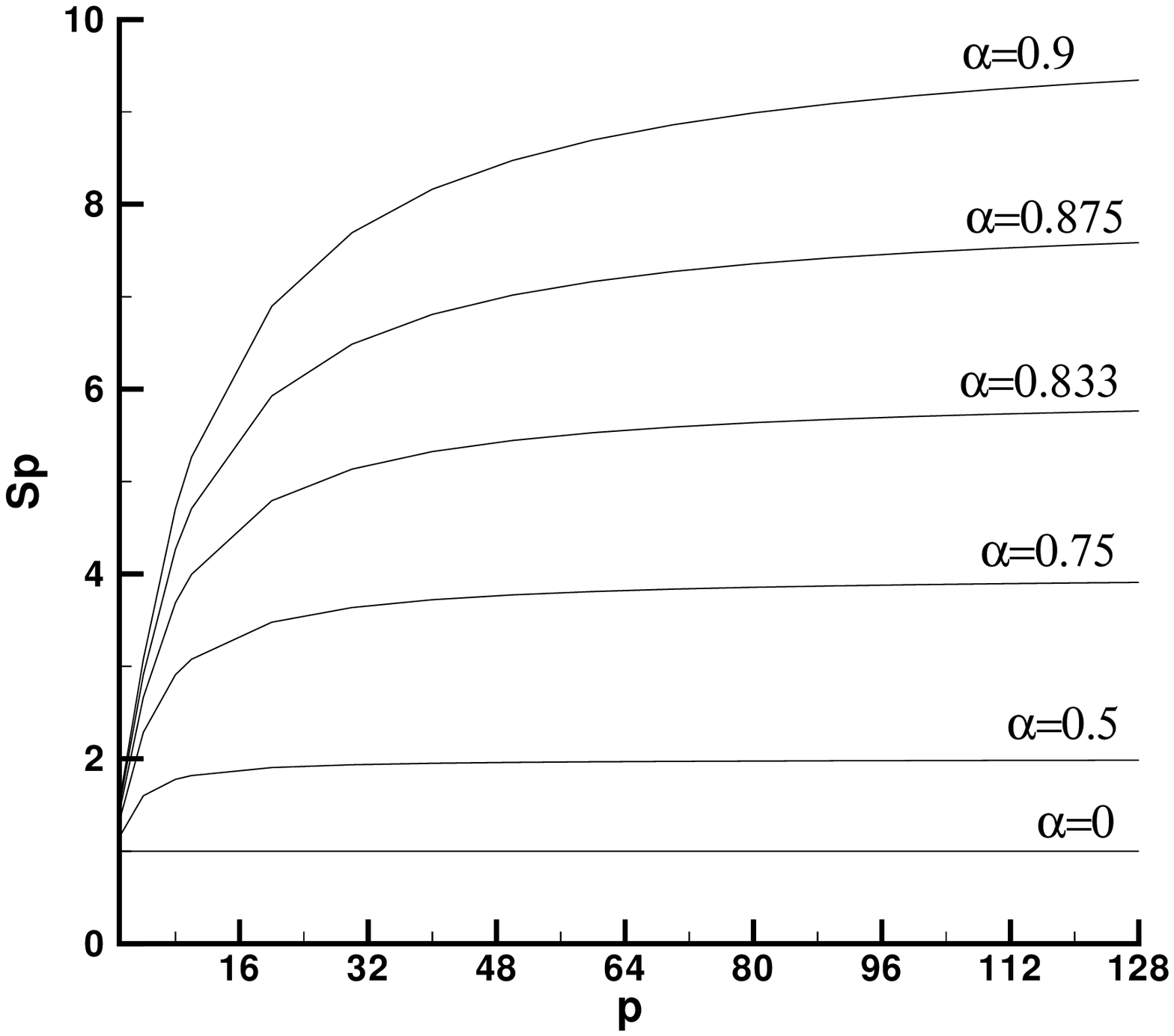}
\end{minipage}%
\begin{minipage}[c]{.5\textwidth}
 \centering
  \includegraphics[width=\textwidth]{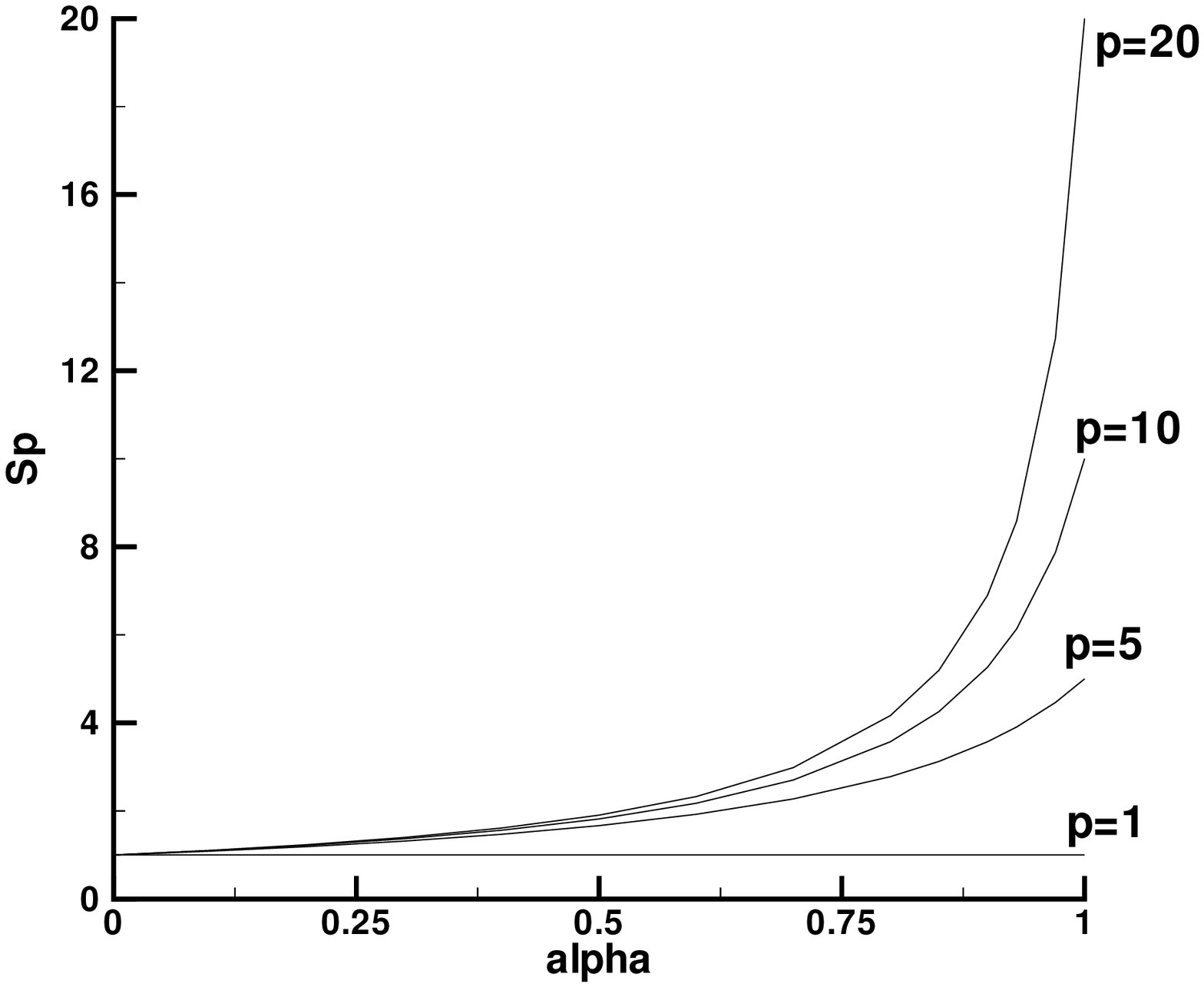}
\end{minipage}
\caption{Speedup $S_p$ as a function of number of processors $p$
for various parallel fractions $\alpha$ (left),
speedup $S_p$ as a function of parallel fraction $\alpha$ for
various number of processors $p$ (right).}
\label{f:sp_alpha}
\end{figure}
\begin{figure}[!ht]
\centering
\begin{minipage}[c]{.4\textwidth}
 \centering
  \caption{Efficiency $E_p$ as a function of parallel fraction $\alpha$
           for various number of processors~$p$}
    \label{f:ep_alpha}
\end{minipage}%
\begin{minipage}[c]{.6\textwidth}
 \centering
   \includegraphics[width=\textwidth]{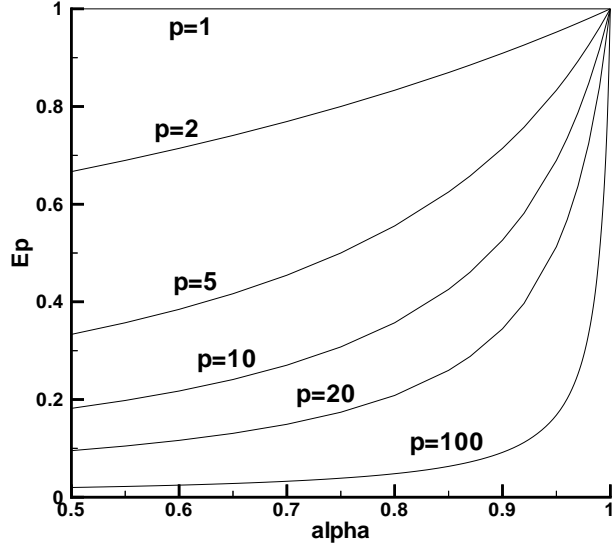}
\end{minipage}
\end{figure}

The value of $T_p$ is given by
\begin{equation}
T_p=[(1-\alpha)+\alpha/p]T_1.
\end{equation}
To get the value of $\alpha$ one may use a profiler. The final
formulas for $S_p$ and $E_p$ are as follows:
\begin{equation}
S_p(p, \alpha)=\frac{p}{p-\alpha(p-1)},
\label{e:psir_sp}
\end{equation}
\begin{equation}
E_p(p, \alpha)=\frac{1}{(1-\alpha)p+\alpha}.
\end{equation}

The formula~(\ref{e:psir_sp}) presents a simple and general function,
called the Amdahl law. According to this law, the speedup upper limit
at $p\rightarrow\infty$ for an algorithm, which has two non-coinciding
parallel and sequential parts, is as follows:
\begin{equation}
S_p(p,\alpha)\le\frac{1}{1-\alpha}.
\end{equation}

To speed up calculations we have to speed up parallel computations,
however, the remaining sequential part slows down the overall
computing process to more and more extent. Even small sequential
fraction may reduce greatly the overall performance.

The figure~\ref{f:sp_alpha} shows the speedup $S_p$ as a function
of number of processors $p$ and parallel fraction $\alpha$.
The efficiency $E_p$ as a functon of $\alpha$ is shown in
the fig.~\ref{f:ep_alpha}. Sequential computations affected speedup
and efficiency particularly  in the region $\alpha>0.9$.
Therefore, even small decrease of sequential computations in
algorithms with high parallel fraction makes speedup and
efficiency abruptly increase (at relatively high $p$).

The PSIR algorithm is coarse-grained and has high efficiency and great 
degree of parallelism comparing to any other parallel algorithm of DSMC of 
unsteady flows for the number of processor $p\leq n$. The maximum value 
of speedup for this algorithm can be obtained at $p=n$. The potential of 
speedup which gives the computer is surplus for $p>n$. Thus, the 
PSIR algorithm for DSMC of unsteady flows has the following range of
efficient usage: $n\gg1$ and $n\geq p$. The value of parallel fraction
$\alpha$ can be very high (up to $0.99\div0.999$) for typical problems of  
molecular gasdynamics~\cite{bykov}. The corresponding speedup is 
$100\div1000$. To get the efficiency $E_p\ge0.5$ at $n=100\div1000$ 
it is necessary to have  $p=100\div1000$ respectively.

\subsection[Data Parallelization of DSMC]%
{Data Parallelization (DP) of DSMC~\cite{grishin}}

The computing time of each DSMC problem is determined by the inner loop (1)
time. The duration of this loop depends on the number of particles in the 
\begin{figure}[!ht]
\centering
\includegraphics[scale=0.6]{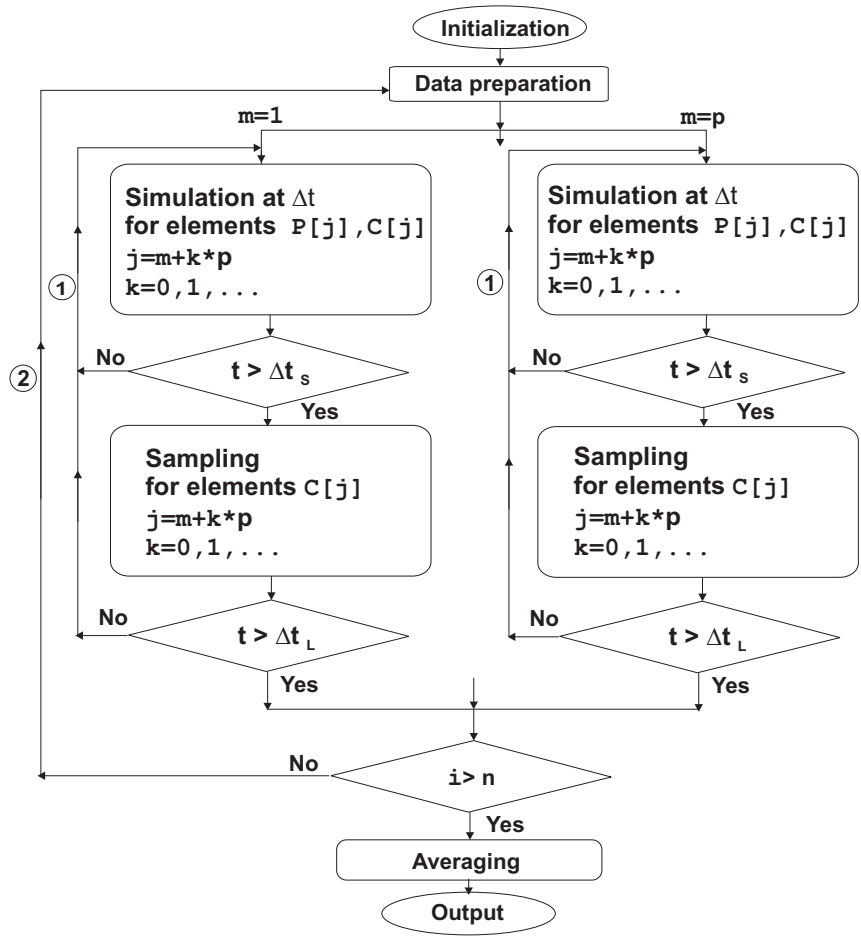}
\caption{General flowchart of DP algorithm;
$j$ --- index of data element}
\label{f:dp_g}
\end{figure}
domain and the number of cells. It was stated above that the inner loop 
consists of  two consecutive stages. The data inside each stage are 
independent. The elements \verb+P[k]+ are processed at the first stage, 
whereas the elements \verb+C[k]+ --- at the second one (the elements of 
arrays~\verb+P+ and \verb+C+ are mutually independent). Since the 
operations on each of these elements are independent it is possible to
process them parallel. Each processor takes elements from 
particle array~\verb+P+ and cell array~\verb+C+ according to its unique
ID-number, i.e. the m-th processor takes the $m$-th, ($m+p$)-th, 
($m+2p$)-th, etc. elements, where ``$m$'' is the processor ID-number.
This rule of particle selection provides good load balancing 
because various particles require different time to process and
they are located randomly in the array~\verb+P+ .
\begin{figure}[!ht]
\centering
\includegraphics[scale=0.5]{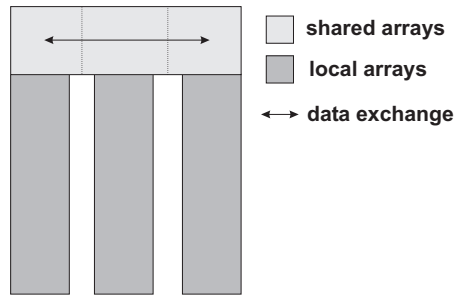}
\caption{Scheme of memory usage for DP algorithm (three processors)}
\label{f:dp_mem}
\end{figure}

The synchronization of processors is performed before the next loop iteration 
starts. Before the second stage begins it is necessary to fill the readdressing 
array \verb+LCR+. The complete information about the array~\verb+P+ is 
required for readdresing procedure. This task can not be parallelized, so it is 
performed by one processor. There are two synchronization points before the 
readdressing and after the one. The reduction of the computational time is 
due to the decrease of the amount of data which has to be processed by each 
processor  ($N_p/p$ and $N_c/p$ instead of $N_p$ and $N_c$). After the 
inner loop is passed the processors also need to get synchronized.
The figure~\ref{f:dp_g} shows the general flowchart of DP algorithm.

The data from the array~\verb+P+ is required to perform the operations on  
elements of array~\verb+C+. This data is located in the array~\verb+P+  
randomly. These arrays are stored in the shared memory in order to reduce 
the large data exchange between processors. The memory conflicts (several 
processors read the same array element) are excluded by the algorithm. The 
semaphore technique is used for processors synchronization.
The scheme of memory usage is depicted in the fig.~\ref{f:dp_mem}.

\section{Algorithm of Two-Level Parallelization with Static Load Balancing}
\label{sec:tlp}
It was stated above that the potential of the multiprocessor system is surplus 
for the realization of the PSIR algorithm when the required number of 
\begin{figure}
\centering
\includegraphics[scale=0.6]{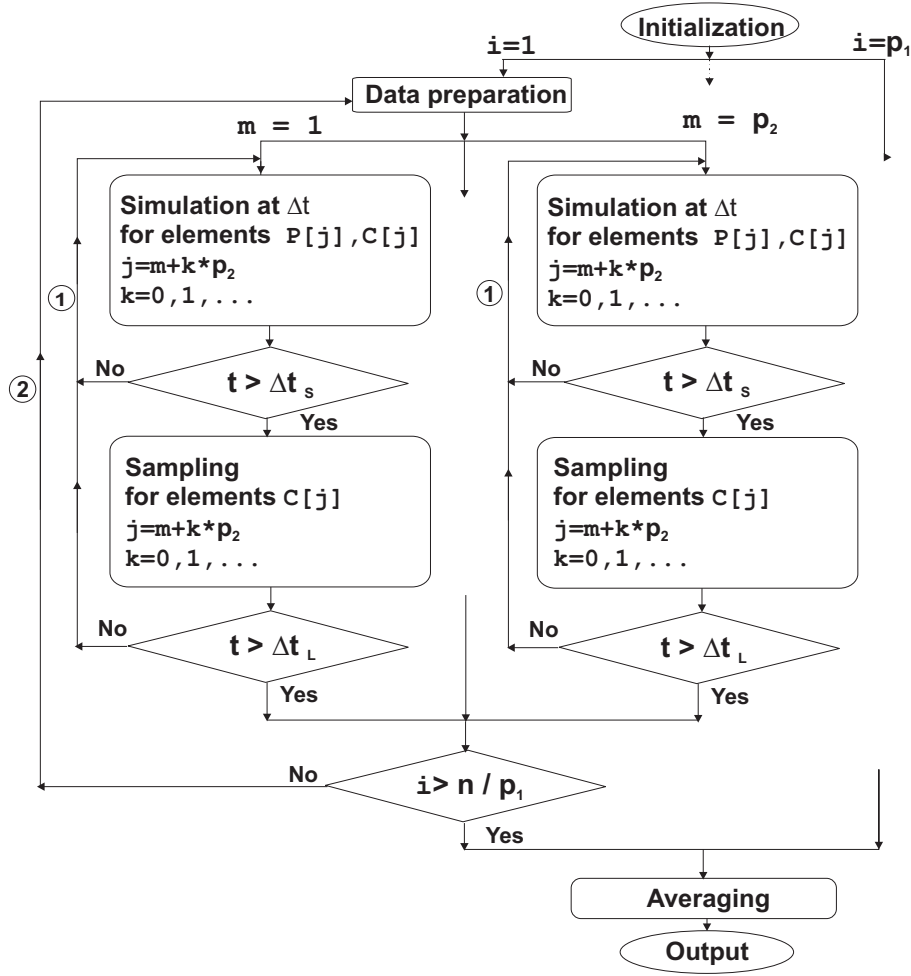}
\caption{Algorithm of two-level parallelization.
$\Delta t$ --- time step,
$\Delta t_s$ --- interval between samples,
$\Delta t_L$ --- time of a single run,
$t$ --- current time,
$i$ --- run number (first level),
$p_1$ --- number of first level processors,
$m$ --- second level processor ID-number,
$p_2$ --- number of second level processors,
$j$ --- index of array element.
}
\label{f:general_tlp}
\end{figure}
statistically independent runs $n$ is significantly less than the number of 
processors $p$ ($n\ll p$). In this case the efficient usage of computer 
resources of $p$-processor system can be provided by the implementation 
of an algorithm of two-level parallelization (TLP algorithm). The general 
flowchart of TLP algorithm is shown in the fig.~\ref{f:general_tlp}. The first 
level of parallelization corresponds to the PSIR algorithm, the data 
parallelization is employed for the second level inside each independent run. 
The TLP algorithm is a parallel algorithm with static load balancing.
\begin{figure}
\centering
\includegraphics[scale=0.5]{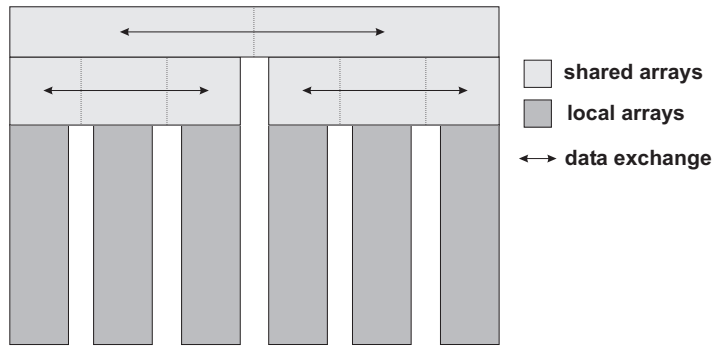}
\caption{Scheme of memory usage for TLP algorithm ($p_1=2$, $p_2=3$)}
\label{f:tlp_mem}
\end{figure}
\begin{figure}[!ht]
\centering
\includegraphics[scale=0.7]{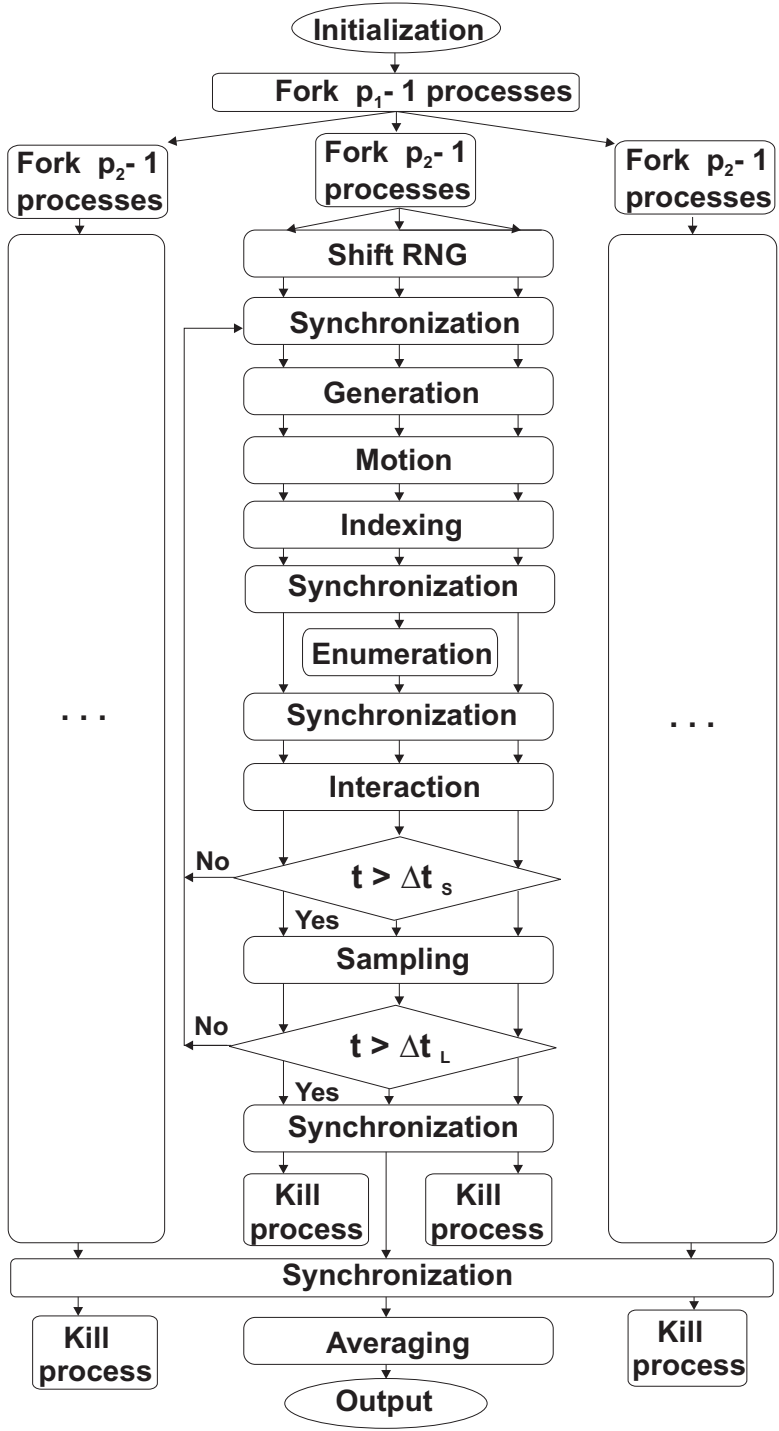}
\caption{Flowchart of TLP algorithm}
\label{f:tlp_scheme}
\end{figure}

The scheme of memory usage for TLP algorithm is depicted
in the fig.~\ref{f:tlp_mem}.
This algorithm requires the memory size to be proportional to the number of 
the first level processors which compute single runs (just the same as for the 
PSIR algorithm). It also requires the arrays for each run to be stored in the 
shared memory as for the data parallelization algorithm in order to reduce the 
data exchange time between processors.

The speedup and the efficiency of the TLP algorithm are governed by the 
following equations:
\begin{eqnarray}
S_p&=&S_{p_1}\cdot S_{p_2}=\frac{p_1}{p_1-\alpha_1(p_1-1)}\,
\frac{p_2}{p_2-\alpha_2(p_2-1)}, \label{e:tlp_sp}\\
E_p&=&E_{p_1}\cdot E_{p_2}=\frac{S_p}{p_1\cdot p_2}, \label{e:tlp_ep}
\end{eqnarray}
where indices `1' and `2' correspond to parameters on the first level and 
on the second one.

The figure~\ref{f:tlp_scheme} shows the detailed flowchart of TLP algorithm 
for unsteady flow simulation. There are five synchronization points in the 
algorithm. The four of them correspond to the DP algorithm. The last 
synchronization has to be done after termination of all runs. The 
synchronization is employed with the aid of the semaphore technique. In
this version the iterations of the outer loop (2) are fully distributed between
the first level processors. This algorithm requires $n$ to be multiply by $p$ 
for uniform distribution of computer resources between single runs. In order 
to make the runs statistically independent we have to shift the random 
number generator in each run.

The HP/Convex Exemplar SPP-1600 system with
8 processors, 2Gb of memory and peak performance
1600 Mflops was used for algorithm test.

To simulate the conditions of a single user in the system we measured the 
execution time of the parent process which makes the start-up initialization 
before forking child processes and data processing after passing parallel code  
(this process has the maximum execution time).
\begin{figure}[!ht]
\centering
\includegraphics[scale=0.5]{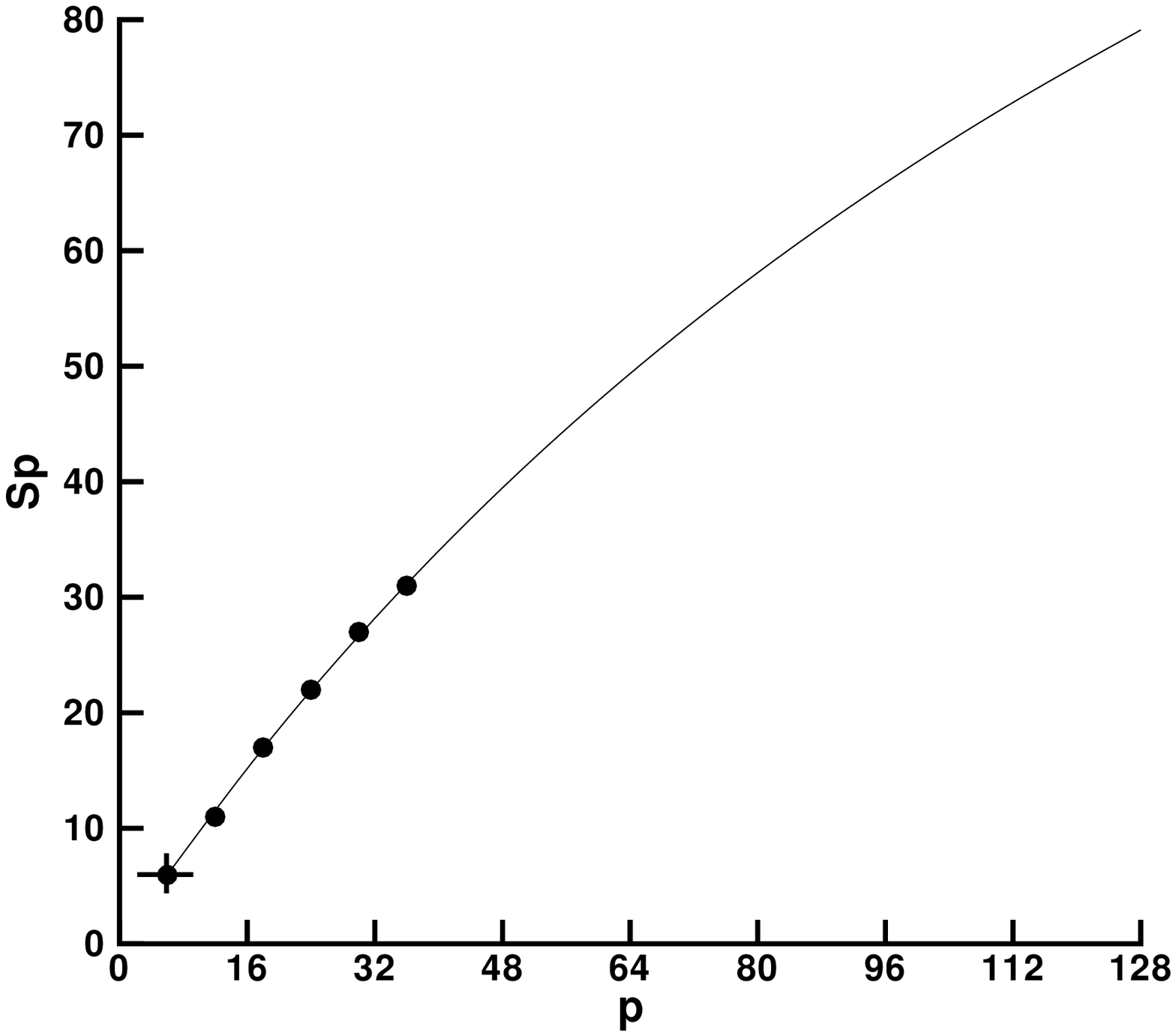}
\includegraphics[scale=0.5]{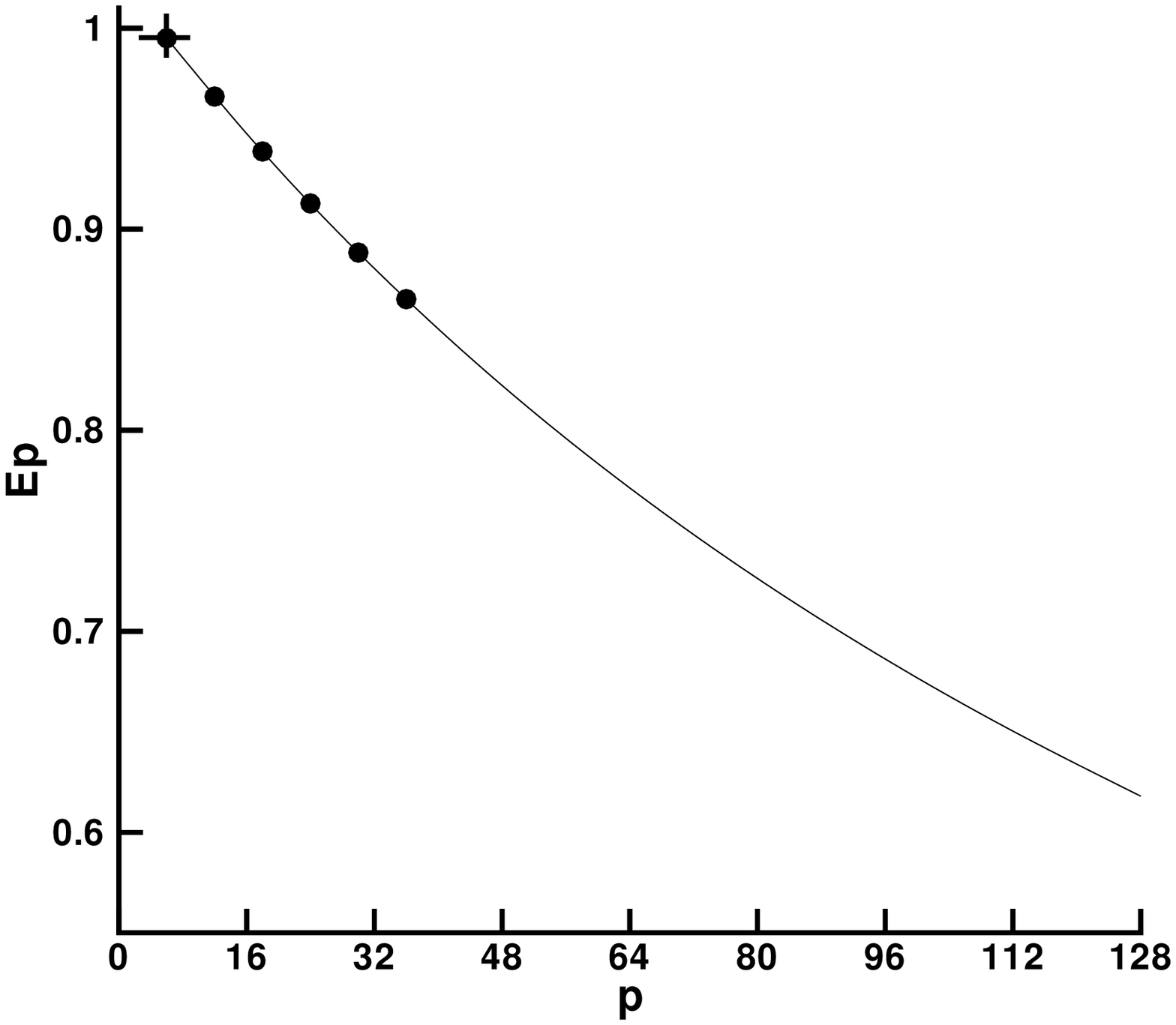}
\caption{Speedup $S_p$ (top) and efficiency $E_p$ (bottom) of TLP algorithm
(circles --- experiment,
curve --- formulae(\ref{e:tlp_sp},\ref{e:tlp_ep}),
cross --- speedup of PSIR algorithm)
}
\label{f:tlp_sp}
\end{figure}

The amount of parallel and sequential code was obtained from the program 
profiling data using standard \verb+cxpa+ utility.

The simulation of unsteady 3-D water vapor flow in the inner
atmosphere of a comet was carried out in order to study the speedup 
and the efficiency which yields this algorithm. The number of  the first level 
processors $p_1$ was fixed and equal to 6. The number of the second level 
processors $p_2$ was varied from 1 to 6. The value of parallel fraction
$\alpha_1$ and $\alpha_2$ were 0.998 and 0.97 respectively. The 
figure~\ref{f:tlp_sp} depicts the experimental results (circles) and 
theoretical curves for speedup and efficiency as functions of the total number
of processors $p=p_1\cdot p_2$. The same figure shows the value (marked by 
cross-sign) of speedup and efficiency of  the PSIR algorithm (TLP algorithm 
turns into PSIR algorithm at $p_2=1$).

Thus, the TLP algorithm gives the possibility to significantly reduce the 
computational time required for the DSMC of unsteady flows using 
multiprocessor computers with shared memory. The range of the efficient 
usage of this algorithm is the condition $n\ll p$. Moreover,  the number of 
processors  $p$ has to be multiply by $n$ in order to provide good load 
balancing.

\section[Algorithm of Two-Level Parallelization with Dynamic Load Balancing]%
{Algorithm of Two-Level Parallelization\\ with Dynamic Load Balancing}

The TLP algorithm with static load balancing described in 
section~\ref{sec:tlp} has several drawbacks. It does not provide good load 
balancing (hence, one may get low efficiency) in the following cases:
\begin{enumerate}
\item the ratio $p/p_1$ is not integer (part of  processors are not used);
\item each run has non-parallelized code with total sequential fraction
equal to $\beta_\ast$, which depends on the starting sequential fraction
$\beta=1-\alpha$ and the number of processors~$p_2$:
\end{enumerate}
\begin{equation}
\beta_\ast=\frac{\beta}{\beta+\frac{1-\beta}{p_2}}.
\label{e:beta_ast}
\end{equation}
At small values of $\alpha$ or large values of $p_2$
some processors may be idle in each run. This leads to
non-efficient usage of computer resources for high values of $p_1$.

\begin{figure}[!ht]
\centering
\includegraphics[scale=0.6]{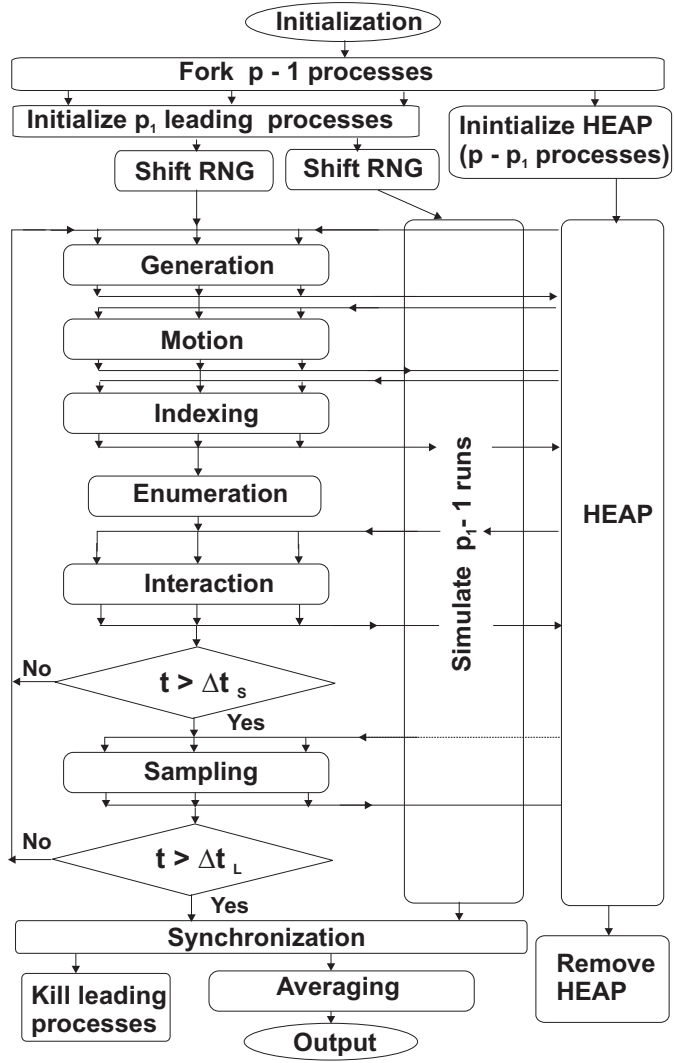}
\caption{Flowchart of TLPDPR algorithm}
\label{f:tlpdpr_scheme}
\end{figure}

The increase in efficiency can be obtained by usage of dynamic load 
balancing with the aid of dynamic processor reallocation (DPR). The idea of 
the algorithm is as follows. Let us conditionally divide all available 
processors into two parts: leading processors $p_1$ and supporting 
processors which form the so called ``heap'' (the number of heap-processors 
is $p-p_1$). Each leading processor is responsible for its own run. This 
algorithm is similar to that of TLP but here there is no hard link of 
heap-processors with the specific run. Each leading processor reserves the 
required number of heap-processors before starting parallel computations
(according to a special allocation algorithm). After exiting from parallel 
procedure the leading processor releases allocated heap-processors. This 
algorithm makes it possible to use idle processors more efficiently, in fact this 
leads to execution of parallel code with the aid of more processors than in 
the case of TLP algorithm with static load balancing. The flowchart of 
TLPDPR algorithm is presented in the fig.~\ref{f:tlpdpr_scheme}.

The speedup which yields this algorithm is determined by the following basic 
parameters: the total available number of processors in the system $p$, the 
required number of independent runs $p_1=n$ ($p_1\ll p$), the sequential 
fraction of computational work in each run $\beta$ and the algorithm 
of heap-processors allocation. In this paper we use the following allocation 
algorithm:
\begin{equation}
p_2^\prime=(1+\PRI)p_2,\quad
\PRI=0\ldots\PRI^\ast,
\label{e:dpr}
\end{equation}
where $p_2^\prime$ --- the actual number of the second level processors, 
$\PRI$ --- the parameter which is estimated by experimental results
of similar problems, 
$\PRI^\ast$ --- the estimated upper limit of the efficient range of parameter 
$\PRI$. 
\begin{figure}[!ht]
\centering
\includegraphics[scale=0.6]{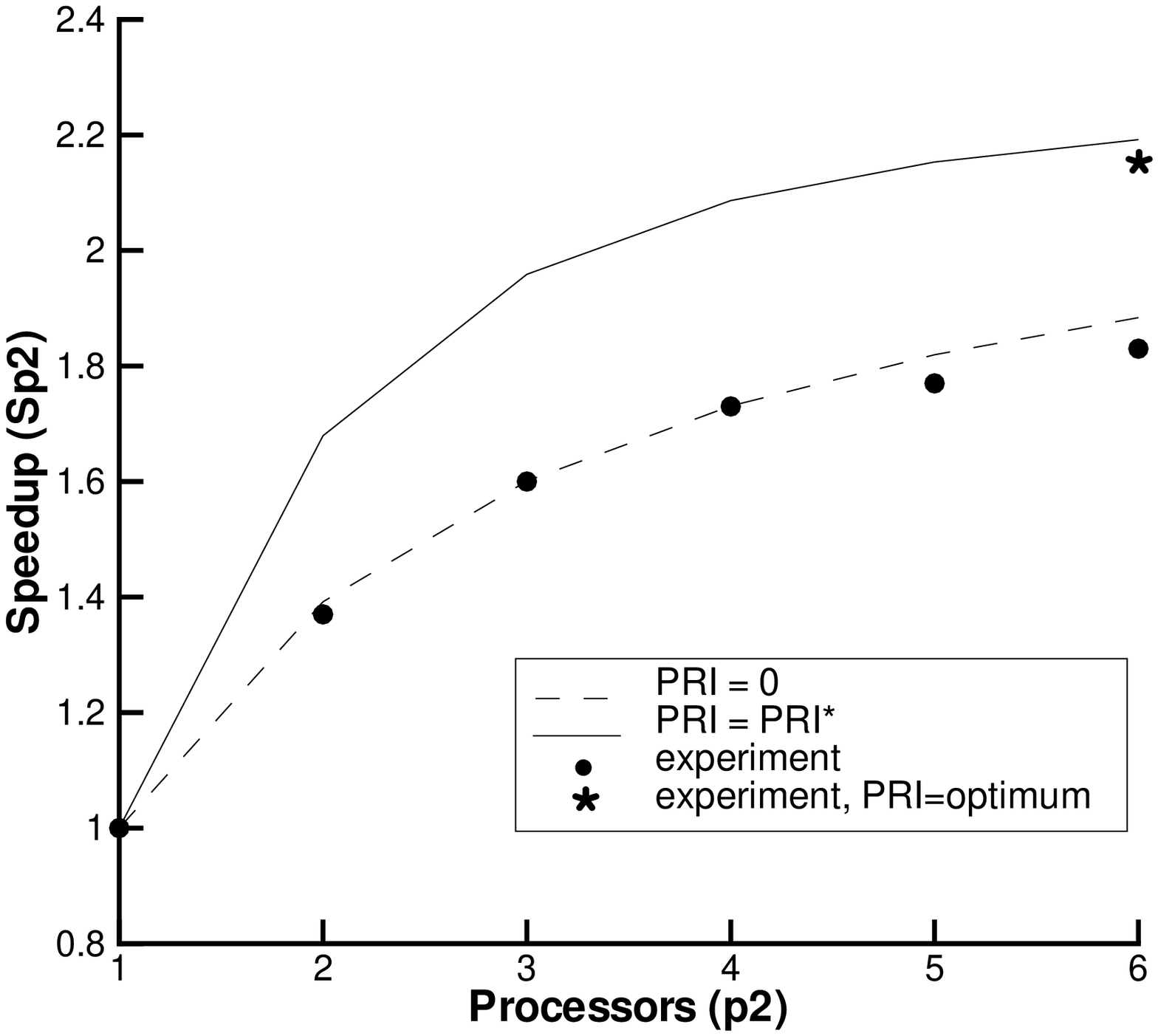}
\caption{Speedup on the second level as a function of 
number of second level processors $p_2$ for algorithms of
TLP ($\PRI=0$, dashed line) and TLPDPR ($\PRI=\PRI^\ast$, solid line),
circles --- experiment ($p_1=6$, $p=36$, $\PRI=0$),
asterisk --- optimal value of parameter $\PRI$
}
\label{f:dpr1}
\end{figure}
\begin{figure}[!ht]
\centering
\includegraphics[scale=0.6]{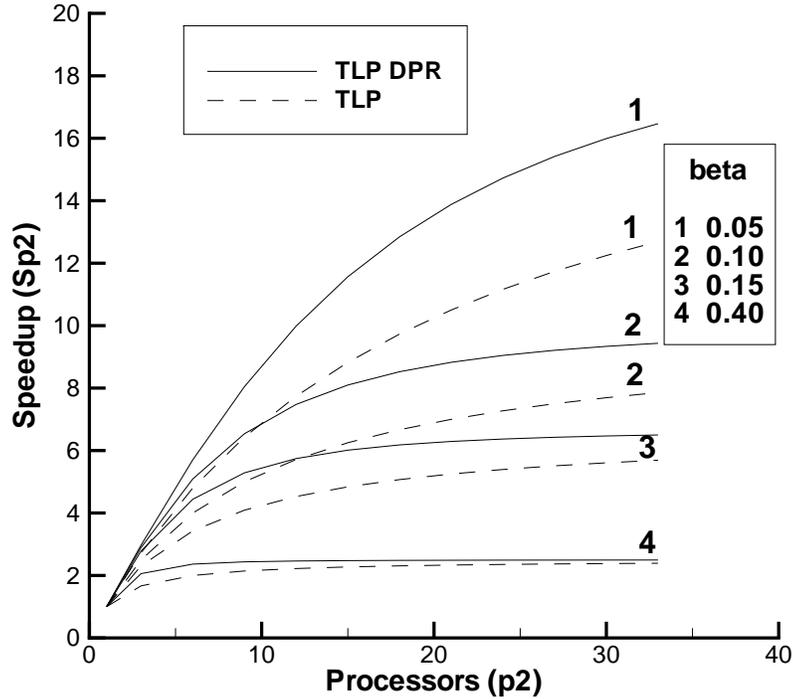}
\caption{Speedups of TLP (dashed line) and TLPDPR (solid line) 
algorithms as functions of number of the second level processors $p_2$ 
for various parallel fractions $\beta$ on the second level}
\label{f:dpr2}
\end{figure}
\begin{figure}[!ht]
\centering
\includegraphics[scale=0.6]{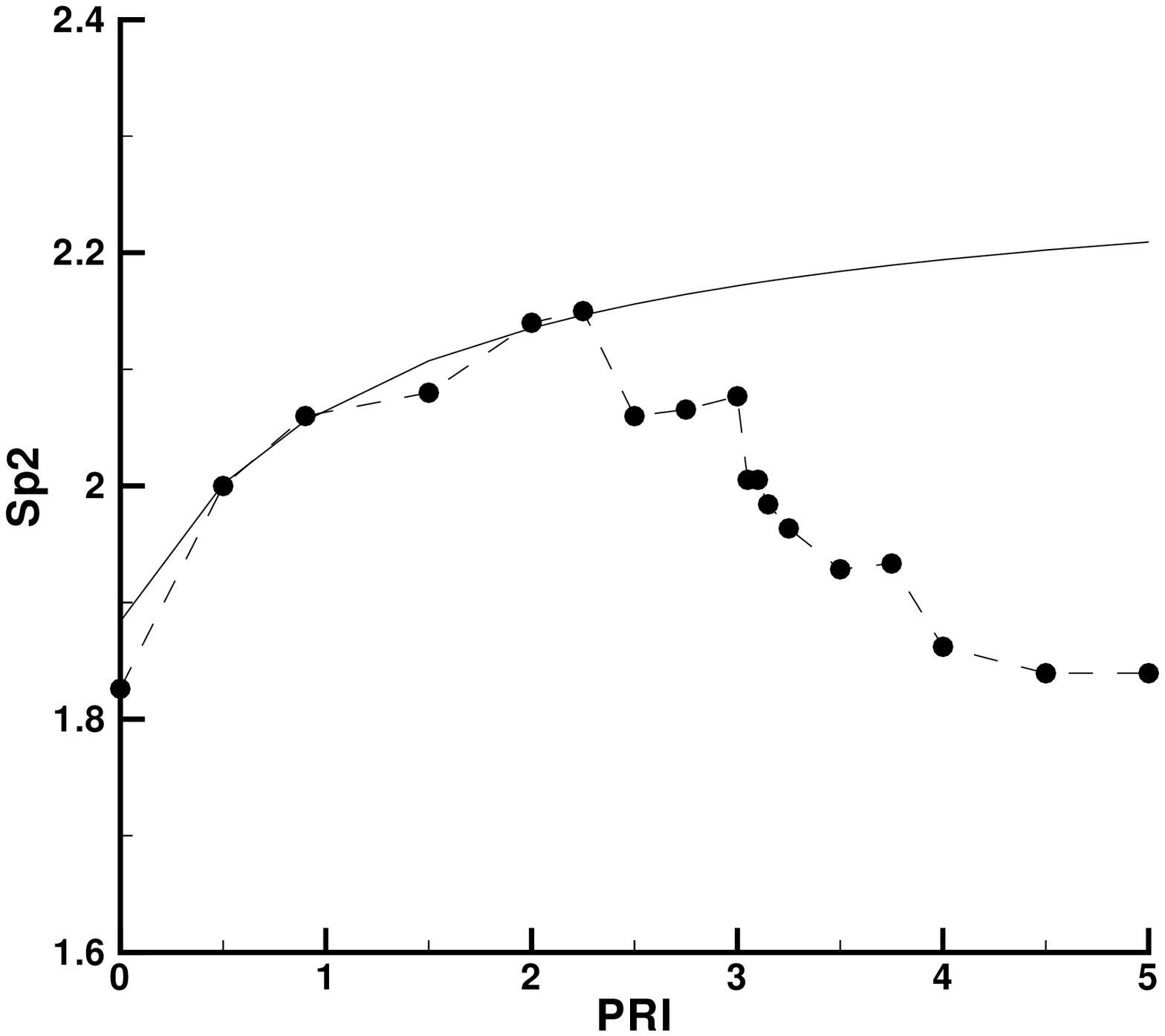}
\caption{Speedup on the second level $S_{p_2}$ as a function of $\PRI$
($p_1=6$, $p=36$,\protect\newline $\PRI=0\ldots\PRI^\ast$).
Solid line --- theory, dashed line --- experiment approximation}
\label{f:pri}
\end{figure}

In case of $p$ being multiply by $p_1$ and the value of $\PRI$ is equal to 0, 
this algorithm turns into TLP algorithm. The speedup on the second level
$S_{p_2}$ is governed by the following equation:
\begin{equation}
S_{p_2}=\frac{1}{\beta+\frac{1-\beta}{p_2(1+\PRI)}}.
\label{e:sp2}
\end{equation}
The case when parameter $\PRI$ exceeds a threshold leads to the decrease
of speedup $S_{p_2}$. This decrease is not governed by~(\ref{e:sp2})
owing to overstating demands made by allocation algorithm on system
resources. As a result, this leads to worse load balancing.
The upper limit of the efficient range of parameter $\PRI$ can
be estimated by the following condition:
\begin{equation}
1+\frac{p-p_1}{(1-\beta_\ast)p_1}=(1+\PRI^\ast)p_2.
\label{e:pri_cond}
\end{equation}
It means that we have to find such a value of parameter $\PRI$
for which there is a uniform distribution of all idle processors
at a given moment among runs which perform parallel computations.
The condition for $\PRI^\ast$ as a function of $\beta$ and $p_2$
can be derived from (\ref{e:beta_ast}) and (\ref{e:pri_cond}):
\begin{equation}
\PRI^\ast=\frac{\beta}{1-\beta}(p_2-1).
\label{e:pri_ast}
\end{equation}
The expressions discussed precedingly are undoubtedly correct for $p_1\gg1$.
The value of $S_{p_2}$ at $\PRI=\PRI^\ast$ gives the upper limit of
speedup for a given problem.

To study the characteristics of TLPDPR algorithm we solve the 
problem on unsteady flow past a body. The value of sequential fraction
$\beta=0.437$, $p_1=6$. The speedup as a function of  $p_2$ 
($p_2=1\ldots6$) for $\PRI=0$ and $\PRI=\PRI^\ast$ is depicted in the 
fig.~\ref{f:dpr1}. The same figure shows the results of  calculation for 
$\PRI=0$, the dot (marked by asterisk) corresponds to the optimal value of 
parameter $\PRI$ for $p_2=6$ ($p=36$). The maximum speedup 
$S_{p_2}$ with a given degree of parallelism ($p_1\rightarrow\infty$),
which can be estimated by the formula~(\ref{e:sp2}), comes to 2.3. The 
TLP algorithm gives the speedup ($\beta=0.437$, $p_1=6$, $p=36$) which 
is 80\% of the maximum value. At optimum value of parameter $\PRI$ the 
TLPDPR algorithm gives 93\% for the same case. This is equivalent to the 
usage of TLP algorithm on a 120-processor computer ($p=120$, $p_1=6$, 
$p_2=20$). The figure~\ref{f:dpr2} shows speedups of TLP and TLPDPR 
algorithms as functions of $p_2$ for various $\beta$.

The essential question one can raise about TLPDPR algorithm usage is  how
to determine the optimal value of parameter $\PRI$ apriori.
The value given by (\ref{e:pri_ast}) determines the upper limit of
efficent range of parameter $\PRI=1\ldots\PRI^\ast$. The study of influence of
parameter $\PRI$ on the speedup is presented in the fig.~\ref{f:pri}
for $p_1=6$, $p_2=6$ ($p=36$).
The formula~(\ref{e:sp2}) gives good approximation of experimental
results for the intial range of parameter $\PRI$. Further, we see the
predicted above decrease of speedup owing to inconsistency of
available and required system resources. The latter can be explained in
the following manner. In (\ref{e:sp2}) it is supposed that
released heap processors are allocated instantly in the other runs.
Actually, these processes are non-coinciding, therefore the
condition~(\ref{e:sp2}) requires a probability coefficient which
is a function of parameters of a problem and a computer. This coefficient
has to determine the probability to meet requirements for system resources
while allocating heap processors.

The great flexibility of this algorithm allows its efficient usage
for calculation of both steady and unsteady problems. In case of
steady-state modeling it is possible to perform an additional
ensemble averaging for smaller number of modeling particles. This can
lead to shorter computation time comparing to DP algorithm.
The implemented TLPDPR algorithm has the following advantages 
comparing to the TLP algorithm with static load balancing:
\begin{itemize}
\item TLPDPR algorithm makes it possible to minimize the latency time of 
processors. It provides better load balancing;
\item Better load balancing make it possible to get higher speedups under the 
same conditions.
\end{itemize}

\newpage
\tableofcontents


\bigskip

\bigskip
\addcontentsline{toc}{section}{References}
\begin{thebibliography}{5}
%
\bibitem{bird}
G.A.Bird.
Molecular Gasdynamics and Direct Simulation of Gas Flows. 
Clarendon Press. Oxford. 1994
%
\bibitem{korolev}
Korolev, M.Ya. Marov, Yu. Skorov, M. Aspnas. 
An Implementation of Monte-Carlo 
weighting method on multiprocessor systems. Reports on 
Computer Science\&Mathematics, Abo Akademy, Ser. A, No. 125, 1991.
%
\bibitem{masahiro}
Ota Masahiro, Tanaka Tetsuya.
On the parallel processing of Direct Simulation Monte 
Carlo method. National Aerospace Lab., Proceedings of the 8th NAL Symposium on 
Aircraft Computational Aerodynamics p. 39-44, Nov 01, 1990.
%
\bibitem{bykov}
N.Y.Bykov, G.A.Lukianov.
Parallel Direct Simulation Monte Carlo of Non-stationary Rarefied Gas 
Flows at the Supercomputers with Parallel Architecture.
St.Petersburg. Institute for High-Performance Computing and Databases. 
Preprint N5-97. 1997.
%
\bibitem{furlani}
Furlani T.R., Lordi J.A. 
A Comparison of Parallel Algorithms for the Direct 
Simulation Monte Carlo Method II: Application to Exhaust Plumes Flowfields, 
AIAA Paper 89-1167, June 1989.
%
\bibitem{wilmoth}
Wilmoth R.G. 
Adaptive Domain Decomposition for Monte Carlo Simulations on 
Parallel Processor. In. 17th Rarefied Gas Dynamics Symposium. AIAA, July 1990.
%
\bibitem{boyd1}
Boyd I.D., Dietrich S. 
A Scalar Optimized Parallel Implementation of the DSMC 
Method, AIAA Paper 94-0355, January 1994.
%
\bibitem{boyd2}
Boyd I.D., Dietrich S. 
Scalar and Parallel Optimized Implementation of the Direct 
Simulation Monte Carlo Method, J. Comp. Phys., 1996, vol. 126, p. 328-342.
%
\bibitem{ivanov}
Ivanov, M., Markelov, G., Tylor, S., Watts J. 
Parallel DSMC Strategies for 3D 
Computations, Parallel CFD'96, P. Schiano et al. eds., North Holland, Amsterdam, 
1997, pp. 485-492.
%
\bibitem{robinson1}
Robinson C.D., Harvey J.K. 
The Development of an Efficient Direct Simulation 
Monte Carlo Computation Scheme for Gas Flows in a Parallel Environment. In 
Proceedings of the Fourth International Parallel Computing Workshop, Imperial 
College/Fujitsu Parallel Computing Research Centre, Imperial College, London, 
1995. Copies available from author.
%
\bibitem{robinson2}
Robinson C.D., Harvey J.K. 
Adaptive Domain Decomposition for Unstructured 
Meshes Applied to the Direct Simulation Monte Carlo Method. In J. Periaux, P. 
Schiano, A. Ecer and N. Satofuka, editors, Proceedings Parallel CFD 96. Elsever, 
1996.
%
\bibitem{oh}
Oh C.K., Sinkovis R.S., Cybyk B.Z., Oran E.S.,Boris J.P. 
Parallelization of Direct 
Simulation Monte Carlo Method Combined with Monotonic Lagrangian Grid, AIAA 
J. vol. 34, N7, July 1996, p. 1363.
%
\bibitem{grishin}
I.A.Grishin, V.V.Zakharov, G.A.Lukianov. 
Data Parallelization of Direct Simulation Monte Carlo in Gasdynamics.
St.Petersburg. Institute for High-Performance Computing and Databases. 
Preprint N3-98. 1998.
%
\bibitem{bogd}
A.V.Bogdanov, N.Y.Bykov, G.A.Lukianov.
Distributed and Parallel Direct Simulation Monte Carlo of Rarefied Gas Flows.
Lecture Notes in Computer Science, Vol. 1401. Springer-Verlag, Berlin 
Heidelberg New York (1998)
%
\bibitem{ortega}
J.M.Ortega.
Introduction to Parallel and Vector Solution of Linear Systems.
Plenum Press. New York. 1988
\end{thebibliography}
\end{document}